\pdfoutput 1
\documentclass[12pt,a4paper]{article}
\usepackage{amsmath,amssymb,amsfonts,amsthm,bm,bbm,cancel,wasysym}
\usepackage{epsfig,graphics,graphicx,epstopdf}
\usepackage{array,booktabs,colortbl,colordvi,multirow}
\usepackage{colordvi,color,xcolor}
\usepackage{rotating}

\newcommand{\be}{\begin{equation}} 
\newcommand{\ee}{\end{equation}} 
\newcommand{\bea}{\begin{eqnarray}}  
\newcommand{\eea}{\end{eqnarray}}

\begin{document}


\begin{flushright}
\today\\
CERN-PH-TH-2015-154
\end{flushright}
\vspace*{5mm}

\renewcommand{\thefootnote}{\fnsymbol{footnote}}
\setcounter{footnote}{1}

\begin{center}

{\Large {\bf Diboson resonant production in non-custodial
    composite Higgs models}}\\  
\vspace*{0.75cm}
{\bf A.\ Carmona${}^a$\footnote{E-mail: carmona@itp.phys.ethz.ch},}
{\bf A.\ Delgado${}^{b,c}$\footnote{E-mail: antonio.delgado@nd.edu},} 
{\bf M.\ Quir\'os${}^{d}$\footnote{E-mail: quiros@ifae.es}} and 
{\bf J.\ Santiago${}^{e}$\footnote{E-mail: jsantiago@ugr.es}}

\vspace{0.5cm}

{\it ${}^a$Institute for Theoretical Physics, \\
ETH Zurich, 8093 Zurich, Switzerland}\\
\vspace{0.5cm}
{\it ${}^b$ Department of Physics, University of Notre Dame,\\
Notre Dame, IN 46556, USA}\\
\vspace{0.5cm}
{\it ${}^c$ Theory Division, Physics Department, CERN,\\
CH-1211 Geneva 23, Swizterland}\\
\vspace{0.5cm}
{\it ${}^d$Instituci\'o Catalana de Recerca i Estudis  
Avan\c{c}ats (ICREA) and\\IFAE-UAB 08193 Bellaterra, Barcelona, Spain}\\
\vspace{0.5cm}
{\it ${}^e$Departamento de F\'{\i}sica Te\'orica y del Cosmos and CAFPE,\\
Universidad de Granada, Campus de Fuentenueva, E-18071 Granada, Spain}\\
\vspace{0.5cm}

\end{center}
\vspace{.25cm}

\begin{abstract}
 
\noindent  
We show that the recently reported excess in resonant diboson
production can be explained in the context of non-custodial composite
Higgs models. Dibosons are generated via the s-channel exchange of
massive vector bosons present in these models. We discuss the
compatibility of the signal excess with other diboson experimental 
searches.
We also discuss the tension between diboson production and other
experimental tests of the model that include electroweak precision
data, dilepton, dijet and top pair production
and show that there is a region of parameter space in which
they are all compatible with the excess. 

\end{abstract}

\renewcommand{\thefootnote}{\arabic{footnote}}
\setcounter{footnote}{0}

\newpage

\section{Introduction}
\label{section_Intro}

The ATLAS collaboration has recently reported an intriguing excess in
hadronically decaying diboson ($WW/WZ/ZZ$) resonant production
peaked around $\sim 2$ TeV~\cite{Aad:2015owa}. 
The observed excess corresponds to significances of 3.4 $\sigma$,
2.6 $\sigma$ and 2.9 $\sigma$ for the $WZ$, $WW$ and $ZZ$ channels
respectively. Hadronically decaying $Z$ and $W$ are not easy to
distinguish so the three channels are not exclusive.
As a cuantitative example, for a $W^\prime$ with a 2 TeV mass
decaying to $WZ$, we have the following $95\%$ C.L. expected (exp) and
observed (obs) limits
\begin{equation}
\sigma(pp\to W^\prime \to WZ)< 12\mbox{ fb (exp)}, 
\quad
\sigma(pp\to W^\prime \to WZ)< 36\mbox{ fb (obs)}.
\label{atlas:vv:hadronic}
\end{equation}
Although the excess is not statistical significant yet, it is
interesting to entertain the possibility that it corresponds to a real
signal of new physics. There is a similar 
search for hadronically decaying diboson production in
CMS~\cite{Khachatryan:2014hpa}.
Interestingly enough there is a small excess localized
near the $2$ TeV region. For a $W^\prime$ with a 2 TeV mass decaying
to $WZ$ CMS finds
\begin{equation}
\sigma(pp\to W^\prime \to WZ)< 8\mbox{ fb (exp)},
\quad \sigma(pp\to W^\prime \to WZ)< 13\mbox{ fb (obs)}.
\label{cms:vv:hadronic}
\end{equation}
An interpretation of these excesses in a number of models
has been recently presented in~\cite{Fukano:2015hga}. In this article
we will show that they can be  explained in the context of
non-custodial composite Higgs models without conflicting with any
other current experimental bound.

A natural prediction of composite Higgs models is the presence of
composite vector
resonances that mix with the Standard Model (SM) electroweak bosons.
They typically have a large coupling to the longitudinal components
of the $W$ and $Z$ bosons and therefore their decay into pairs of SM
gauge bosons is usually sizeable.
Thus, they are prime candidates to explain an excess in diboson
resonant production.
The mass scale of the diboson
excess $M\sim 2$ TeV is however smaller than the typical masses of
the vector resonances in composite Higgs models, which, due to a
sizeable tree-level contribution to the $S$ parameter, are expected to
be in the $M\gtrsim 2.5-3$ TeV region~\cite{Agashe:2003zs}. 
However, using a soft-wall
construction of higher-dimensional holographic duals of composite
Higgs models it has been shown that lighter vector resonances are
compatible with electroweak precision tests (EWPT) both in
custodial~\cite{Falkowski:2008fz} and
non-custodial~\cite{Cabrer:2010si} 
set-ups. 
In practice this is done by reducing the
mixing between the heavy vector resonances and the SM gauge bosons and
the coupling between the former and the SM fermions. Reducing these
mixings and couplings allows for lighter resonances compatible with
EWPT but it also reduces their production cross section and their
decay branching ratio into SM vector bosons, making their collider
phenomenology more challenging~\cite{Carmona:2011ib,deBlas:2012qf}.
Hence there is, in principle,  
 a tension in these models between the constraints from EWPT
and the possible size of anomalous diboson production.
For the sake of simplicity we will consider in this article the 
minimal non-custodial composite
Higgs model as a benchmark and study in detail the compatibility of
the observed excess with constraints from EWPT.~\footnote{Custodial
  models have a similar phenomenology and, although the quantitative
  details will differ, the general features that the diboson excess
  can be explained without conflict with any other experimental
  constraint is likely to hold also in these models.}

  Before explaining the excesses in the context of specific models,
however, one should check whether these excesses are compatible with
the results of other diboson resonant
searches or if, on the contrary, they are already excluded by the latter.
We will show in the next section that, although there is a large
number of related searches for diboson resonance production with no
significant excess, the corresponding bounds are compatible with the
number of events needed to explain the observed excess.
  Of course with a specific
model in mind there are 
also other experimental searches that play a relevant role in constraining
the parameters of the model. In particular the most important ones for
non-custodial composite Higgs models are
dijet, dilepton and top pair production.
We will show that there are regions of parameter space in which the observed
excess can be explained without contradicting any of the existing limits.

The rest of the paper is organized as follows. In
Sec.~\ref{experimental:sec} we will summarize the different
experimental searches for exotic resonances. In Sec.~\ref{model} the
non-custodial model will be presented including the EW fit. The
results are presented in Sec.~\ref{results} and finally
Sec.~\ref{conclu} is devoted to our conclusions. 

\section{Experimental status\label{experimental:sec}}

As mentioned above, before trying to explain the observed excess in
terms of a specific model, we have to make sure that the excess is not already
ruled out by other related resonant searches.
There are a number of other diboson (WW/WZ/ZZ) resonant searches performed by
ATLAS and CMS in the semi-leptonic and purely leptonic channels. No
significant excesses are observed in these searches and the resulting
$95\%$ C.L. bounds are reported in the first block of
Table~\ref{bounds}.
\begin{table}[!h]
\begin{center}
\begin{tabular}{|lcr|c|c|c|c|}
	\hline
\multicolumn{3}{|c|}{Channel} & Process & 1.8 TeV&1.9 TeV& 2.0 TeV\\
\hline
ATLAS &$\ell\nu jj$& \cite{Aad:2015ufa}&$pp\to W'\to WZ$& 13 fb& 12 fb & 10 fb\\
CMS &$\ell\nu j j $&\cite{Khachatryan:2014gha} & $pp\to G^*\to WW$& 6 fb& 4 fb & 3 fb\\
ATLAS& $\ell\ell j j $&\cite{Aad:2014xka}& $pp\to W'\to WZ$& 14 fb& 20 fb &20 fb\\
ATLAS &$\ell\ell j j $&\cite{Aad:2014xka}& $pp\to G^*\to ZZ$&6 fb & 7 fb & 7 fb\\
CMS &$\ell\ell j j $&\cite{Khachatryan:2014gha}& $pp\to G^*\to ZZ$&14 fb & 12 fb & 8 fb\\
ATLAS& $3\ell\nu$&\cite{Aad:2014pha}& $pp\to W'\to WZ$&21 fb & 22 fb & 21 fb\\
CMS& $3\ell\nu$ &\cite{Khachatryan:2014xja}& $pp\to W'\to WZ$& 27 fb& 20 fb & 20 fb\\
\hline
ATLAS &$ZH$ &\cite{Aad:2015yza}& $pp\to Z' \to ZH$& 14 fb & 16 fb & $-$\\
ATLAS &$WH$ &\cite{Aad:2015yza}&$pp\to W'\to WH$&31 fb & 37 fb & $-$\\
 CMS &$ZH$ &\cite{Khachatryan:2015bma}& $pp\to Z' \to ZH$& 13 fb & 9 fb & 7 fb\\
CMS &$WH$ &\cite{Khachatryan:2015bma}&$pp\to W'\to WH$&14 fb & 9 fb & 7 fb\\
\hline
ATLAS& $\ell\ell$&\cite{Aad:2014cka}& $pp\to Z'\to \ell \ell$& 0.23 fb& 0.22 fb & 0.20 fb\\
ATLAS& $\ell\nu$&\cite{ATLAS:2014wra}& $pp\to W'\to \ell\nu$&0.54 fb &0.48 fb & 0.44 fb\\
CMS &$\ell\ell$&\cite{Chatrchyan:2012oaa}& $pp\to Z'\to \ell \ell$&0.24 fb & 0.24 fb & 0.24 fb\\
CMS &$\ell\nu$&\cite{Khachatryan:2014tva}& $pp\to W'\to \ell\nu$&0.40 fb & 0.34 fb & 0.30 fb\\
ATLAS& $t\bar{t}$&\cite{Aad:2015fna}& $pp\to Z'\to t\bar{t}$&64 fb &60 fb & 52 fb\\
CMS &$t\bar{t}$&\cite{Khachatryan:2015sma}& $pp\to Z'\to t\bar{t} $&17 fb &14 fb & 11 fb\\
ATLAS& dijets& \cite{Aad:2014aqa}& $pp\to W'\to jj $&270 fb & 184 fb & 119 fb\\
CMS &dijets& \cite{Khachatryan:2015sja}& $pp\to W'\to jj $&205 fb &155 fb & 95 fb\\
\hline
\end{tabular}
\end{center}
\caption{Summary of  the relevant $95\%$ C.L. observed bounds.}
\label{bounds}
\end{table}

At first sight it might seem that these bounds rule out the reported 
excess by ATLAS (and to a lesser extent CMS) in
Eqs.~(\ref{atlas:vv:hadronic}) and (\ref{cms:vv:hadronic}). 
However, a
more detailed look at the results in Ref.~\cite{Aad:2015owa} shows
that the cross section needed to explain the ATLAS excess is in
principle compatible with all these extra constraints.
The most significant excess is observed in the two $100$ GeV bins with invariant
mass $1850\mbox{ GeV}\leq m_{VV^\prime} \leq 2050\mbox{ GeV}$ and it
corresponds to 8, 6.5 and 6.5 events over the expected background for
the $WZ$, $WW$ and $ZZ$ selection regions, respectively~\cite{twiki}. Assuming that
the diboson resonance corresponds to a vector boson exchanged in the
s-channel, the total
number of events can be computed as follows
\begin{equation}
N_{\mathrm{ev}}= \mathcal{L} \times \epsilon \times 
\sigma(p p \to V^\prime) \times
BR(V^\prime \to V_1 V_2) \times BR(V_1 \to jj) \times BR(V_2 \to jj),
\label{nev:signal}
\end{equation}
where $\mathcal{L}=20.3\mbox{ fb}^{-1}$ is the integrated luminosity, 
$V^\prime$ is the intermediate vector boson, $V_{1,2}$ stand
for a $W$ or $Z$ boson and $\epsilon$ is the analysis efficiency.
From Table~1 and Fig.~2~(b) of Ref.~\cite{Aad:2015owa} we
conservatively estimate 
\begin{equation}
\epsilon \approx 0.14 \times 0.7 \approx 0.1,
\end{equation}
where $0.7$ is the fraction of events in the region around the
$V^\prime$ mass.
Inserting the numbers in Eq.~(\ref{nev:signal}) we obtain
\begin{equation}
\sigma(p p \to V^\prime) \times BR(V^\prime \to V_1 V_2) 
\approx
N_{\mathrm{ev}} \mbox{ fb} \approx 6-8 \mbox{ fb},
\end{equation}
where we have assumed similar efficiencies for the reconstruction of
the hadronic $Z$ and $W$ and for simplicity we have neglected the
small difference in their hadronic branching fractions. 
A proper statistical combination of all three different channels has
not been provided by the ATLAS collaboration but they emphasize that a
number of events will populate several regions. Instead of trying to
estimate the contamination across channels we will simply add the
cross sections in all of them. This is a reasonable
approximation, taking into account that we have been conservative when
estimating the efficiencies.
Thus, we see that there is no obvious contradiction between the
required cross section to explain the observed ATLAS excess with the
limits from other searches. The only possible exception is the 
CMS $\ell\nu j j$ analysis~\cite{Khachatryan:2014gha}, which puts a
constraint on  $\sigma(pp\to G^*\to WW)\leq  6-3$ fb for $M=1.8-2$
TeV. As we will discuss below, our model does not give a
contribution to 
this signal with the same strength as to the fully hadronic
one. Furthermore, the interpretation of the bound in terms of a spin-2
resonance makes it difficult to apply it to our model without further
assumptions. 

We finally show in the second and third blocks of Table~\ref{bounds}
other experimental 
constraints that will be relevant for the non-custodial composite
Higgs model that we will use to explain the diboson excess.
Searches for exotic resonances decaying into a W or Z boson and a
Higgs boson have been performed by ATLAS and CMS. No significant
excess is found in these channels and the $95\%$ C.L. bounds on
cross-sections are summarized in the second block of
Table~\ref{bounds}. Finally other resonant channels, as $\ell\ell$,
$\ell \nu$, $t\bar t$ and dijets have been searched for by ATLAS and
CMS and the resulting $95\%$ bounds on cross-sections are summarized
in the third block of Table~\ref{bounds}.  
We will include all the bounds from the second and third blocks of
Table~\ref{bounds} in our numerical analysis below.

\section{Non-custodial composite Higgs models\label{model}}

Composite Higgs models provide an elegant solution to the hierarchy
problem. In these models the Higgs boson is a composite resonance of a
new strongly coupled interaction and its mass is protected by its
finite size. Further assuming that the Higgs is a
pseudo-Nambu-Goldstone boson of a global symmetry of the strongly
interacting sector ensures that its mass is naturally much smaller
than the composite scale~\cite{Kaplan:1983fs}. A common prediction of
these models is the presence of electroweak composite vector
resonances with 
masses around or above the scale of compositeness and a large mixing
with the SM gauge bosons.
Due to a sizeable tree-level contribution to the $S$ parameter, these
vector resonances are however
expected to 
be in the $M\gtrsim 2.5-3$ TeV region, even in custodially symmetric
set-ups~\cite{Agashe:2003zs}. 
Using a soft-wall
construction of higher-dimensional holographic duals of composite
Higgs models it has been shown that lighter vector resonances are
compatible with EWPT both in
custodial~\cite{Falkowski:2008fz} and
non-custodial~\cite{Cabrer:2010si} models. 
Indeed, vector resonances as
light as $\sim 1$ TeV are allowed 
without violating the very stringent constraints
from EWPT. In practice this is done by effectively reducing
the mixing between the composite vector resonances and the SM gauge
fields. The $T$ parameter, which is volume enhanced in these models,
is proportional to the square of this mixing whereas the $S$
parameter, which does not have the volume enhancement and is therefore
naturally smaller than the $T$ parameter, is linearly proportional to
the mixing. Thus, the reduction in the mixing naturally brings the
model back in the allowed ellipse in the $S-T$ plane with much lighter
vector resonances. In the explicit construction used
in~\cite{Cabrer:2010si} the
coupling of the composite vectors to the light SM fermions is also
reduced with respect to models in AdS$_5$ thus reducing the production
cross section of the vector resonances and making their collider
phenomenology more difficult to
explore~\cite{Carmona:2011ib,deBlas:2012qf}. The reduced mixing
between the composite and SM vectors also has a direct implication
relevant for this work. This mixing governs the decay of the massive
vector resonances into pairs of SM gauge bosons and therefore reducing
it to ease the constraints from EWPT also reduces the potential signal
to explain the diboson excess.

In order to be quantitative and explore a region of parameter space as
large as possible in this class of models without the burden of
constructing complete non-trivial gravitational backgrounds, 
we consider a simplified
version of the non-custodial composite Higgs model in which we include
the minimal number of fields that are relevant for the diboson
production and for the associated constraints. Specifically we
consider only the first resonances with the quantum numbers of the
electroweak SM gauge bosons. Color octet vector and fermionic
resonances are also typically present in these models but their
features are model dependent and would unnecesarily complicate the
current analysis. Furthermore there are stringent constraints from
flavor, dijet and top production on color octet vector
resonances~\cite{Barcelo:2011wu} that
are therefore expected to be heavier than the needed $\sim 2$ TeV to
explain the diboson excess.

The relevant part of the Lagrangian reads
\begin{equation}
\mathcal{L}= \mathcal{L}_{\mathrm{SM}}^{(0)} +
\mathcal{L}_{\mathrm{Gauge}}^{(1)}
+\mathcal{L}_{\mathrm{Fermion}}^{(1)}+\mathcal{L}_{\mathrm{Higgs}}^{(1)}, 
\end{equation}
with
\begin{eqnarray}
\mathcal{L}_{\rm Gauge}^{(1)}&=&
-\frac{1}{4}W_{\mu\nu}^{(1)i}W^{(1)i\mu\nu}
-\frac{1}{4}B_{\mu\nu}^{(1)}B^{(1)\mu\nu}
+\frac{1}{2}M^2W_{\mu}^{(1)i}W^{(1)i\mu}
+\frac{1}{2}M^2B_{\mu}^{(1)}B^{(1)\mu}\nonumber\\
&&-g\varepsilon^{ijk}
\partial_{[\mu} W_{\nu]}^{(1)i}W_{\mu}^{(1)j}W_{\nu}^{(0)k}
-\frac{1}{2}g\varepsilon^{ijk}
\partial_{[\mu} W_{\nu]}^{(0)i}W_{\mu}^{(1)j}W_{\nu}^{(1)k},
\label{eq:laggauge}
\\
\mathcal{L}^{(1)}_{\mathrm{Higgs}} &=& 
\frac{g g_V}{2} W^{(1)i}_\mu \phi^\dagger
\mathrm{i} \overset{\leftrightarrow}{D}^{i\,\mu} \phi
+\frac{g^\prime g_V}{2} B^{(1)}_\mu \phi^\dagger
\mathrm{i} \overset{\leftrightarrow}{D}^{\mu} \phi,
\label{eq:laghiggs}
\\
\mathcal{L}^{(1)}_{\mathrm{Fermion}} &=& 
\sum_{\psi_L}g g_{\psi_L}
W^{(1)i}_\mu \bar{\psi}_L
\gamma^\mu \frac{\sigma^i}{2}  \psi_L
+\sum_\psi g^\prime g_\psi Y_\psi B^{(1)}_\mu \bar{\psi} \gamma^\mu \psi,
\label{eq:lagferm} 
\end{eqnarray}
where 
$\overset{\leftrightarrow}{D}^{i\,\mu}\equiv \sigma^i D^\mu -
\overset{\leftarrow}{D}^\mu \sigma^i$,
$\overset{\leftrightarrow}{D}^{\mu}\equiv D^\mu -
\overset{\leftarrow}{D}^\mu$, $Y_\psi$ stands for the hypercharge of
fermion $\psi$ and $\psi_L$ and $\psi$ run over all left-handed SM
fermions and all SM fermions, respectively. $D^\mu$ stands for the SM
covariant derivative and the square brackets denote antisymmetrization
with respect to the indices.
We assume that all quarks, except for the right-handed
top and the left-handed top-bottom doublet, have the same
coupling, while all leptons share the same coupling, which can be different
in general from the one of quarks.~\footnote{In composite Higgs models
  this can be naturally explained due to the different mass generation
  mechanism between quarks and leptons~\cite{delAguila:2010vg}.}
Thus we have five relevant parameters that control the mixing between
the composite vectors and the SM gauge bosons ($g_V$) and the couplings of the
composite vectors to the SM leptons ($g_{l}$), light quarks ($g_q$),
left-handed top 
and bottom quarks ($g_{q_3}$) and right-handed top quark ($g_{t_R}$),
respectively (the mass of the massive vectors is essentially fixed by
the scale of the ATLAS excess).
All these parameters but $g_{t_R}$ enter the contributions to
EWPT. $g_q$ controls the production cross section of the composite
resonances and $g_V$ their decay into pairs of SM gauge
bosons. $g_{t_R}$ has a global impact on the decays of the composite
vectors because it is expected to be sizeable (from the top mass) and
therefore can give a large contribution to the total width of the
composite vectors.

The Lagrangian (\ref{eq:laghiggs}) contains mixing terms between the
composite $Z$ and $W^{\pm}$ resonances and their SM counterparts,
leading to the following mass terms 
\begin{eqnarray}
\mathcal{L}\supset\frac{1}{2}(Z_{\mu}^{(0)}~Z^{(1)}_{\mu})\mathcal{M}_Z^2
\begin{pmatrix}Z^{(0)}_{\mu}\\Z^{(1)}_{\mu}\end{pmatrix}
+(W_{\mu}^{(0)+}~W_{\mu}^{(1)+})\mathcal{M}_W^2
\begin{pmatrix}W^{(0)-}_{\mu}\\W^{(1)-}_{\mu}\end{pmatrix},
	\label{eq:massmat}
\end{eqnarray}
with 
\begin{eqnarray}
\mathcal{M}_{X}^2\approx \begin{pmatrix}m_X^2 &g_{V}m_X^2
\\ g_Vm_X^2&M^2\end{pmatrix},  \quad \mathrm{for~} X=Z,W,
\end{eqnarray}
where $W^{(1)\pm}_\mu$ and $Z_{\mu}^{(1)}$ (and $A_\mu^{(1)}$) 
are defined exactly as their SM
counterparts.
Even though there are no interaction terms in (\ref{eq:laggauge})
between one heavy resonance and two light states (reminiscent of the
orthonormality of the different KK modes in the holographic duals), 
the diagonalization of
(\ref{eq:massmat}) leads to such couplings when the required rotation
is implemented in the $V_1^{(0)}V_2^{(0)}V_3^{(0)}$ and the
$V^{(1)}_1V^{(1)}_2 V_3^{(0)}$ couplings. In particular, after going
to the physical basis $(Z_{\mu},Z^{\prime}_{\mu})$ and
$(W_{\mu}^{\pm},W_{\mu}^{\prime \pm})$, one gets
$\mathcal{O}(m_W^2/M^2)$ suppressed  couplings 
$Z^{\prime} W^{+} W^-$, $W^{\prime\pm }W^{\mp} Z$ and $A^{\prime} W^{+}W^{-}$ (where
$A^{\prime}$ is just $A^{(1)}$, since the photon resonance does not
mix with the zero-mode). Note that in the unitary gauge that we are
working on the couplings of the heavy vector resonances to the SM
gauge bosons are suppressed. Nevertheless the enhanced decay into the
longitudinal components of the SM gauge bosons provides the sizeable
branching ratio that we would expect from the couplings to the
electroweak Goldstone bosons in a renormalizable gauge. 
On the other hand, since the couplings
$Z^{(1)}Z^{(0)}H$ and $W^{(1)\pm}W^{(0)\mp}H$ are already present in
(\ref{eq:laghiggs}), the effect of the aforementioned rotation is
subleading and we can safely neglect it for the present
phenomenological study.  

In our model SM diboson resonances proceed through the following
processes
\begin{eqnarray}
q\bar{q}^\prime &\to& W^{\prime\, \pm} \to W^\pm Z,
\\
q\bar{q} &\to& Z^\prime/A^\prime \to W^+ W^-.
\end{eqnarray}
We therefore have no anomalous $ZZ$ production and will have to
account for the observed excess with the $WW$ and $WZ$ samples.
In the fully hadronic case we will neglect the small differences in
the reconstruction of the hadronic $W$ and $Z$ and therefore we will
simply add the $WW$ and $WZ$ production cross sections in order to
estimate the signal excess.
These processes are accompanied by the corresponding ones with the
Higgs boson
\begin{eqnarray}
q\bar{q}^\prime &\to& W^{\prime\, \pm} \to W^\pm H,
\\
q\bar{q} &\to& Z^\prime \to Z H.
\end{eqnarray}

As we have indicated, the heavy vector boson production is controlled by
the coupling to light quarks $g_q$. Their decay into dibosons
(including the Higgs boson) is proportional to $g_V$ but it is also
sensitive to all the other couplings through the vectors' widths.
They can also decay in any of the other possible channels, namely into
dijets (through $g_q$), tops (through $g_{q_3},g_{t_R}$) 
or dileptons (through $g_{l}$). As we
have discussed in Sec.~\ref{experimental:sec}, all
these channels have quite stringent constraints on the production
cross section of new resonances with a $\sim 2$ TeV mass.
Regarding the very stringent bound from the
CMS $\ell\nu j j$ search~\cite{Khachatryan:2014gha}, our model
contributes to it through both $WW$ and $WZ$ production. Nevertheless,
the $WZ$ channel contributes with a factor $1/2$ with respect to the
$WW$ one, due to the multiplicity of the latter in the semi-leptonic
decay. Thus, neglecting the differences in efficiencies (and hadronic
decays) in the hadronic $W$ and $Z$ reconstructions we have
\begin{equation}
\sigma(p p \to Z^\prime/A^\prime \to W W) 
+\frac{1}{2}\sigma(p p \to W^\prime \to W Z)  \lesssim 6-3~\mbox{fb,
($M=1.8-2$ TeV).}
\end{equation}
We have checked that this bound indeed introduces some tension with
the expected signal in our model, typically leaving a very narrow allowed
strip in the $g_l-g_q$ plane. 
However, this bound requires the very strong assumption that the
efficiencies for a spin-2 mediator are similar to those of a spin-1
mediator like the ones present in our model. Thus, we can only take
this bound as an indication of the possible constraints of the
semileptonic search in the parameter space of our model, and encourage
the experimental collaborations to interpret their limits in the
context of spin-1 resonances.
Neglecting this search, the next most constraining one from the first
block of Table~\ref{bounds} is the ATLAS $\ell \nu jj$ bound. In our
case, we have contribution from both the $WW$ and $ZW$ channels, the
former contributing with a factor of 2 due to the multiplicity. Thus,
we also impose the following constraint on our signal
\begin{equation}
2\sigma(p p \to Z^\prime/A^\prime \to W W) 
+\sigma(p p \to W^\prime \to W Z)  \leq 13-9~\mbox{fb,
 ($M=1.8-2$ TeV).}\label{diboson:upper:bound}
\end{equation}

\begin{figure}
\begin{centering}
\includegraphics[width=0.45\textwidth]{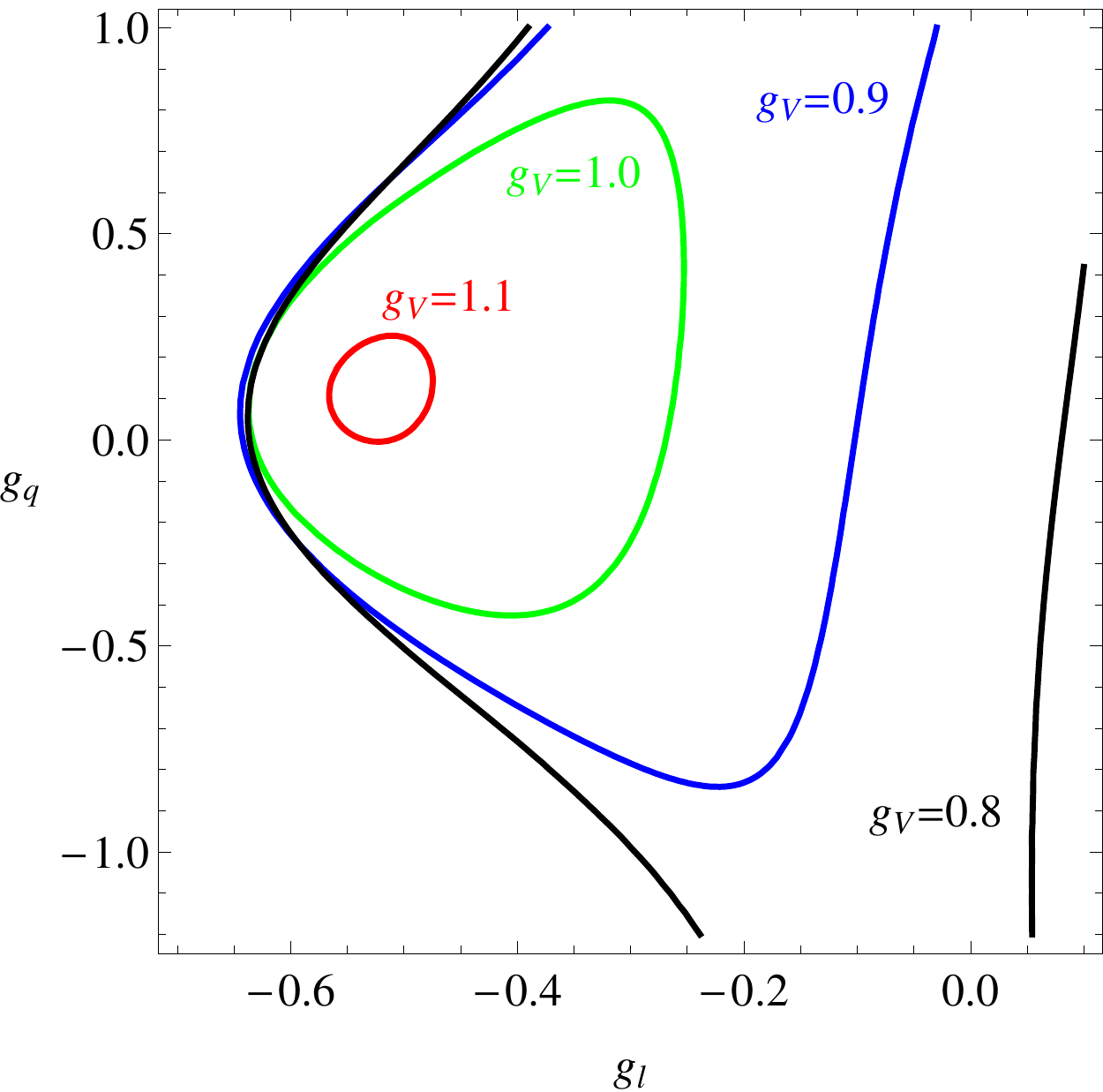}
\hfil
\includegraphics[width=0.45\textwidth]{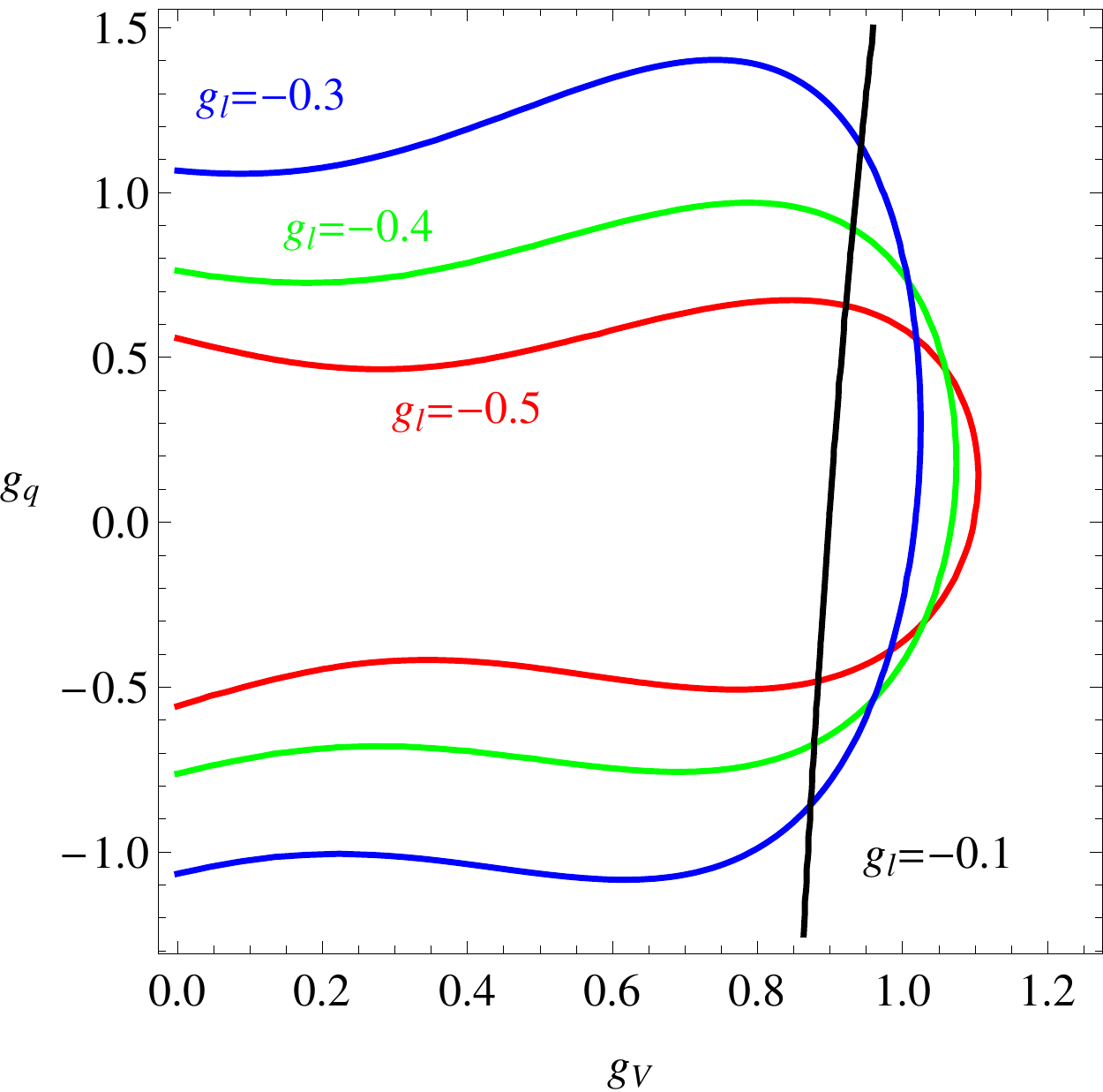}
\end{centering}
\caption{$95 \%$ C.L. allowed regions by EWPT, with $M=2$ TeV and
  $g_{q_3}=0.5$, in the $g_l-g_q$ (left) and $g_V-g_q$ (right) planes.
  The allowed regions are the interior of the corresponding contours
  on the left figure and the region to the left of the contours on the
  right figure.
\label{EWPT:fig}
}
\end{figure}

As we have discussed above, EWPT introduce further constraints in the
parameter space of the model.
In the case of
a universal coupling of light fermions ($g_q=g_l$), the $T$
parameter is proportional to $g_V^2$ whereas the $S$ parameter is
proportional to $g_q g_V$. Thus, in order to have a set of $\sim 2$ TeV  
vector boson resonances compatible with EWPT we need to have
suppressed $g_V$ and $g_q$. In turn this means a supressed production
cross section and supressed branching fraction into dibosons (and
therefore less signal for the diboson excess). In the more general case
$g_q\neq g_l$ that we are considering, EWPT can no longer be parameterized
in terms of oblique parameters but the tension between EWPT and the
generation of our signal is similar. We have implemented EWPT as
follows. First we have computed 
the dimension-6 effective Lagrangian that results after
integrating out the massive vector bosons at tree level. 
The resulting effective Lagrangian
can be read-off from Eqs.~(58-66) of~\cite{Davoudiasl:2009cd} with the
identifications
\begin{eqnarray}
\alpha^N&=& -\frac{g_V^2}{M^2},\quad
\beta_\psi^N= -\frac{g_V g_\psi}{M^2},\quad
\gamma_{\psi \psi^\prime}^N= -\frac{g_\psi g_{\psi^\prime}}{M^2},
\nonumber \\
\alpha^D&=& \beta^D_\psi = \gamma^D_{\psi \psi^\prime}=0.
\end{eqnarray}
We have then used the results
of~\cite{Efrati:2015eaa} 
to implement the constraints at the $Z$-pole and the ones
on~\cite{deBlas:2013qqa} to include the constraints on four-fermion
interactions.  They imply stringent bounds on the parameters of our
model (except for $g_{t_R}$). As an example we show in
Fig.~\ref{EWPT:fig} the allowed region in the $g_l-g_q$ (left) and
$g_V -g_q$ (right) planes for $M=2$ TeV, $g_{q_3}=0.5$ and
different values of the remaining free parameter. As we can 
see from the figure, there is a very stringent constraint on $g_V$
\begin{equation}
g_V \lesssim 1.2\mbox{  (EWPT)},
\end{equation}
quite independent of the other parameters. Reducing the mass of the
vector resonances makes these limits more stringent.
We have only included the tree-level contribution
of the vector resonances to EWPT. Loop contributions can be very
significant in non-custodial composite Higgs
models~\cite{Carmona:2011ib} (see 
also~\cite{Anastasiou:2009rv} for the case of important loop contributions in
custodially invariant models). However they are more model-dependent
and in particular they cannot be directly correlated to the diboson
excess we are trying to explain. Specific realizations of our general
parameterization of non-custodial composite Higgs models will have to
include such constraints to determine the viability of complete models
to explain the observed diboson excess.

\section{Results\label{results}}

The discussion in the previous section makes it clear in which region of
parameter space we can expect composite Higgs models 
to explain the ATLAS diboson signal. We
need a sizeable production cross section of the heavy vector
resonances, which implies a sizeable $g_q$, and a large decay
branching fraction into SM vector bosons, thus the largest possible
$g_V$. A sizeable production cross section then means that the
branching fraction into dileptons and $t\bar{t}$ cannot be too large,
otherwise the very stringent bounds on these channels would be
violated. Finally, we cannot just increase $g_q$ arbitrarily without
clashing with dijet or other diboson constraints. 
As we saw in Fig.~\ref{EWPT:fig},
EWPT play a further, somewhat intrincate, role in fixing the allowed
region of parameter space that might explain the observed ATLAS
excess.
\begin{figure}
\centering
\begin{tabular}{cc}
\includegraphics[width=0.4\textwidth]{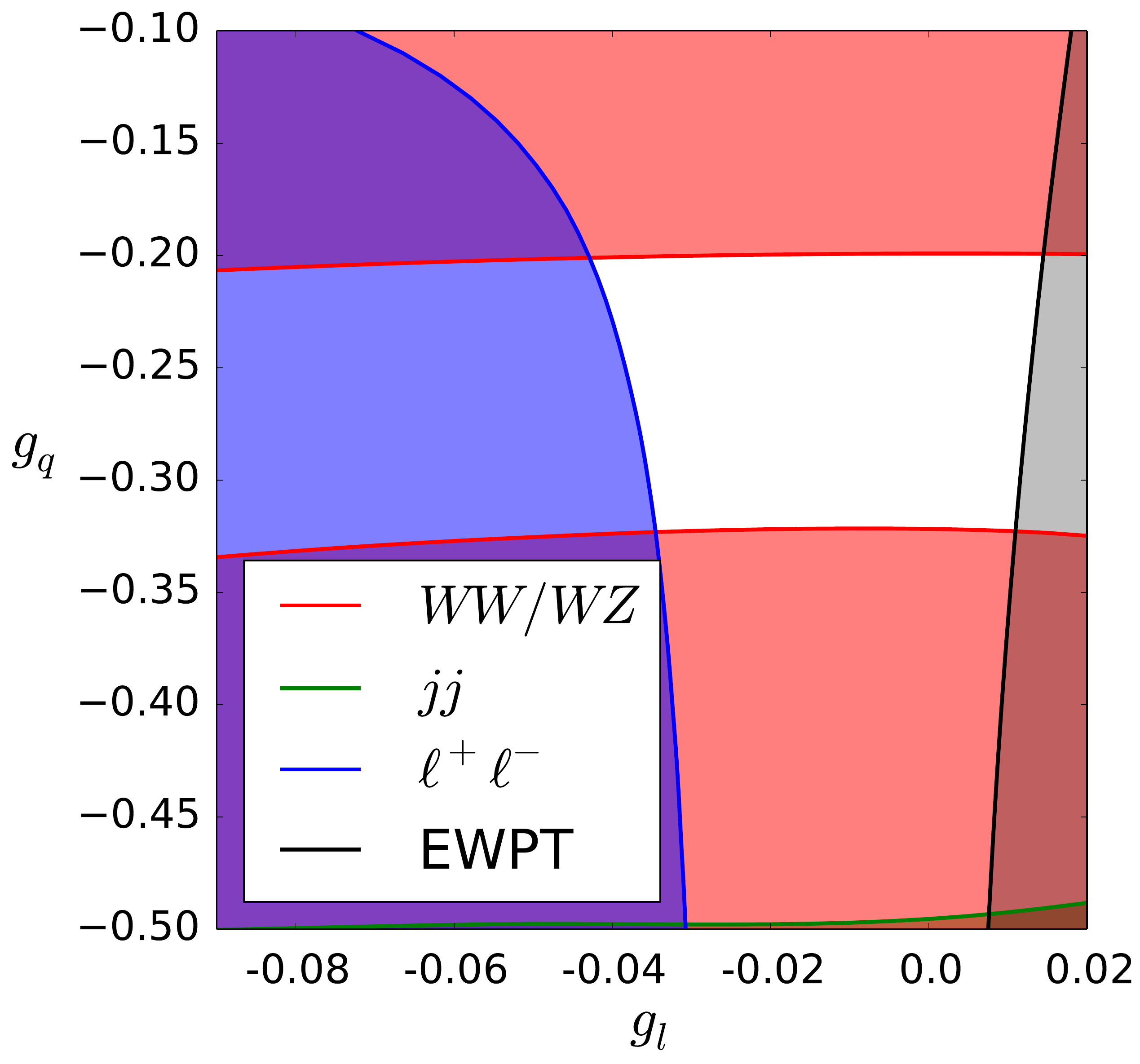}
&
\includegraphics[width=0.4\textwidth]{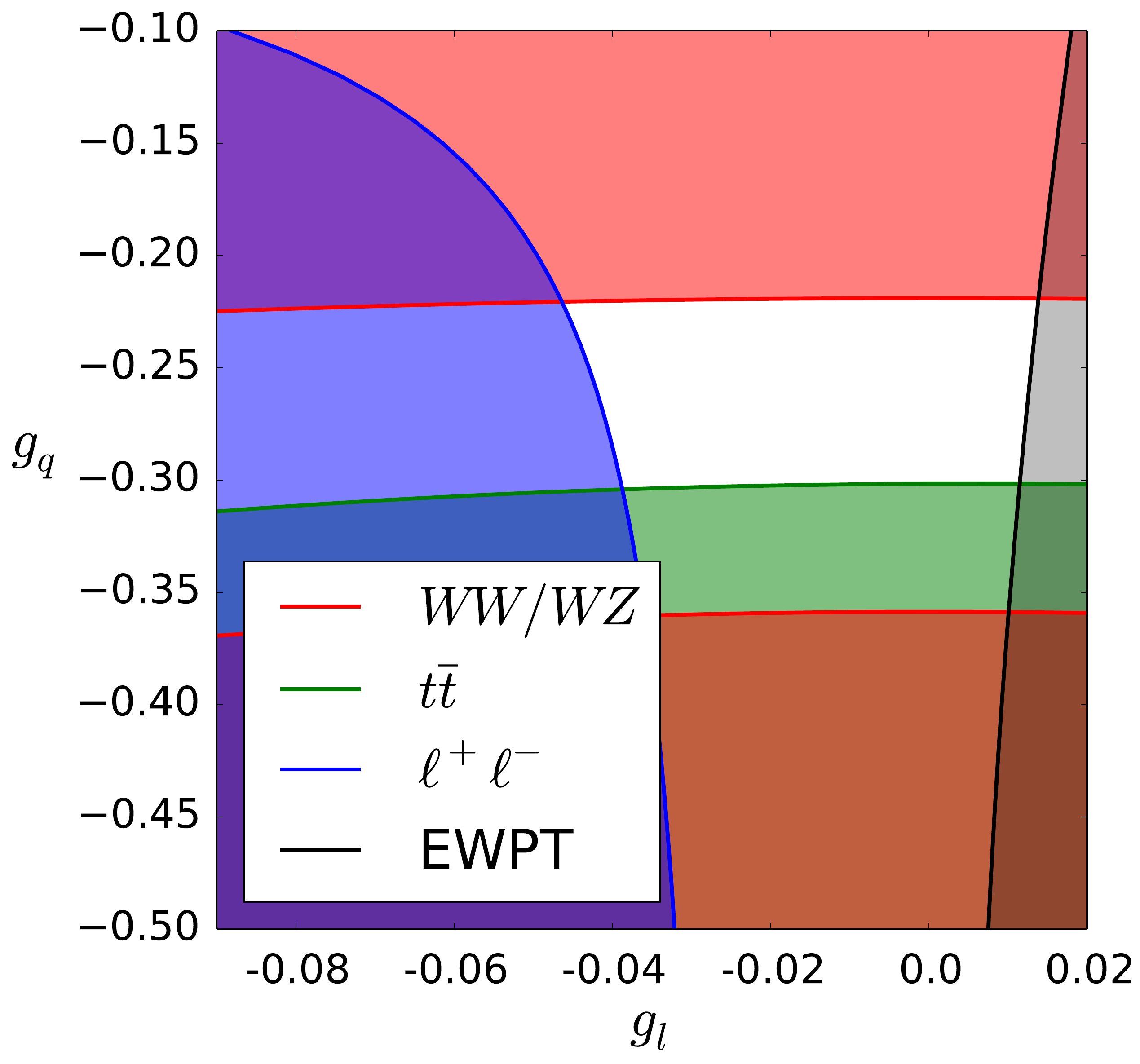}
\\
\includegraphics[width=0.4\textwidth]{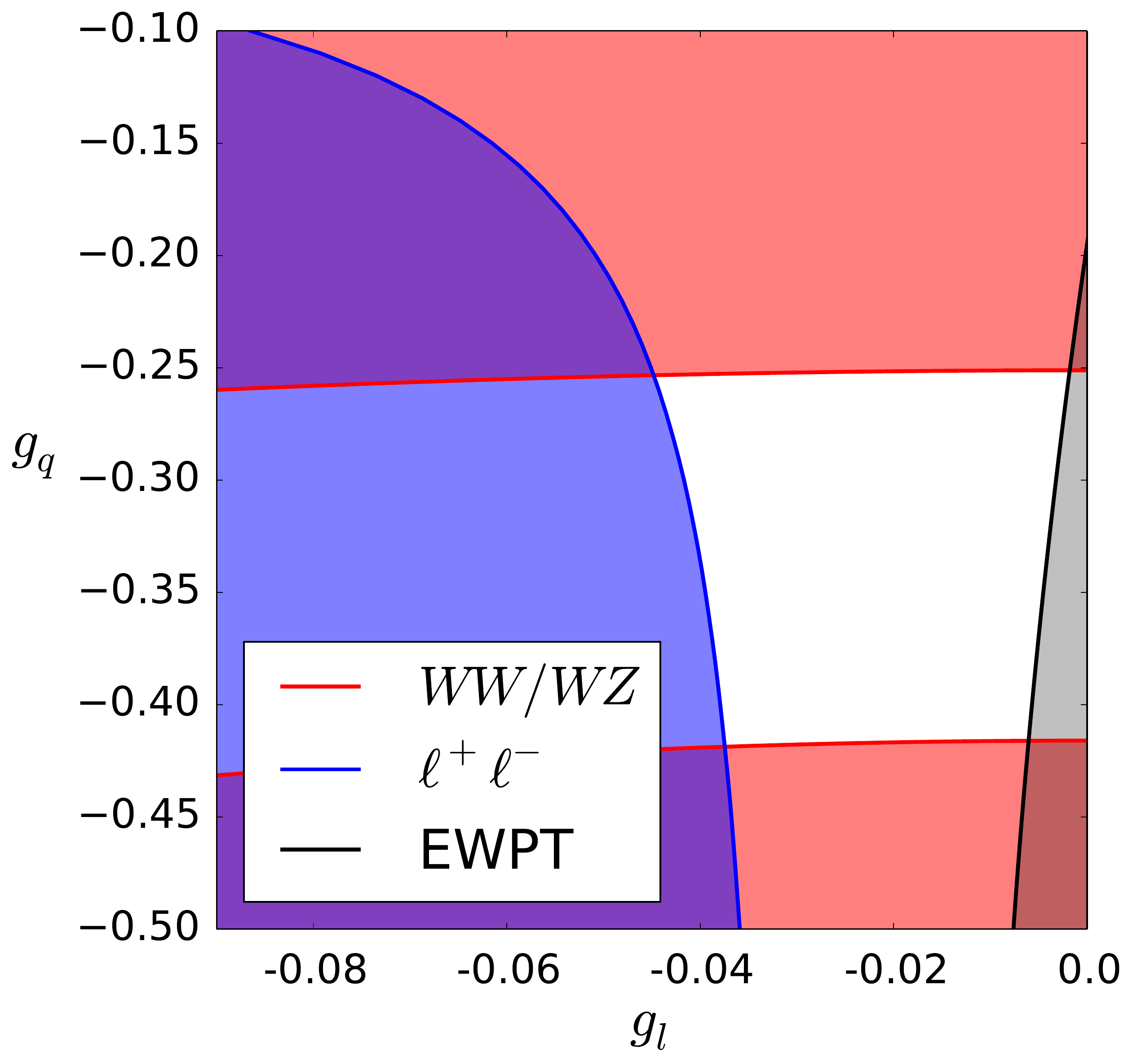}
&
\includegraphics[width=0.4\textwidth]{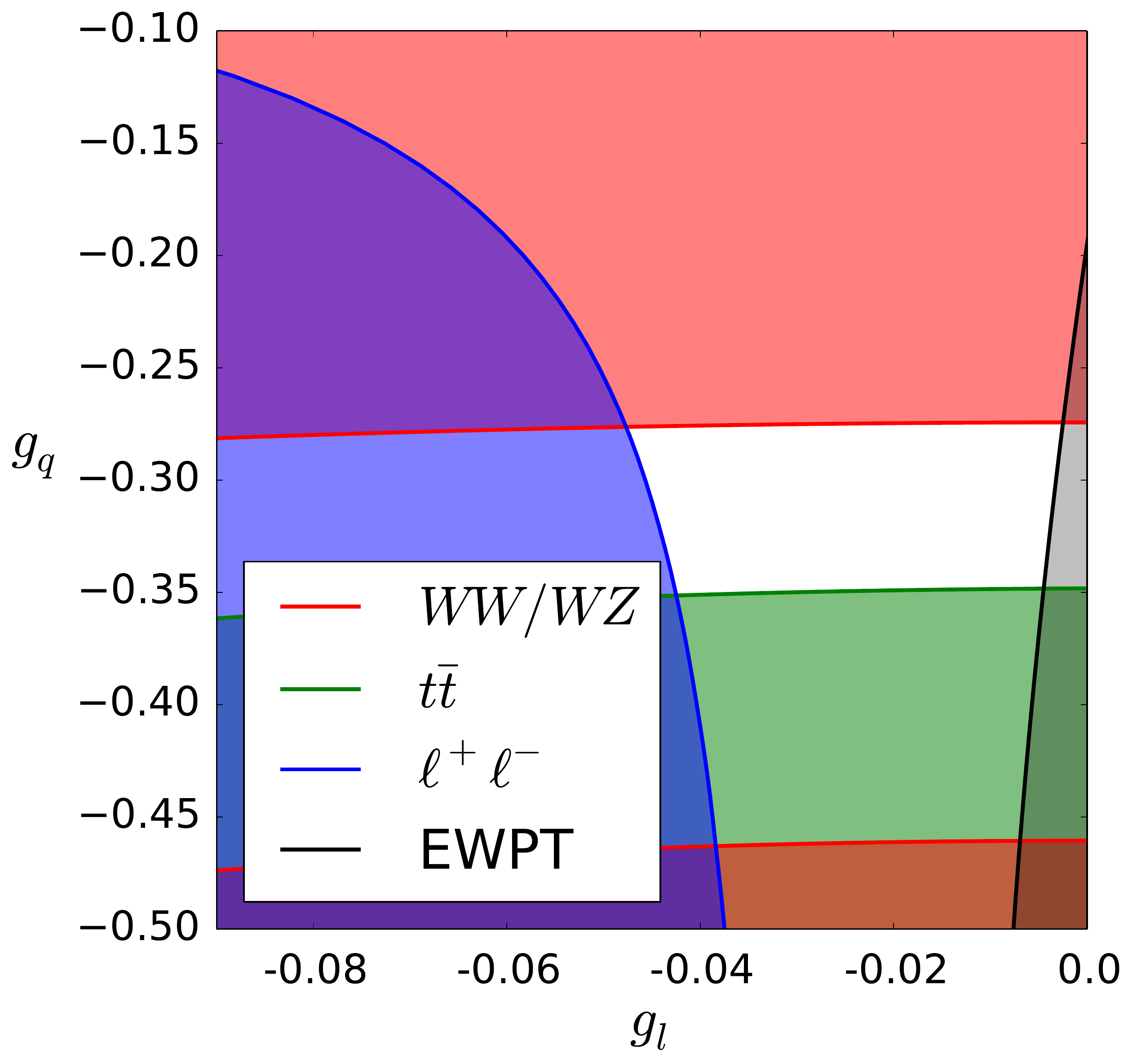}
\\
\includegraphics[width=0.4\textwidth]{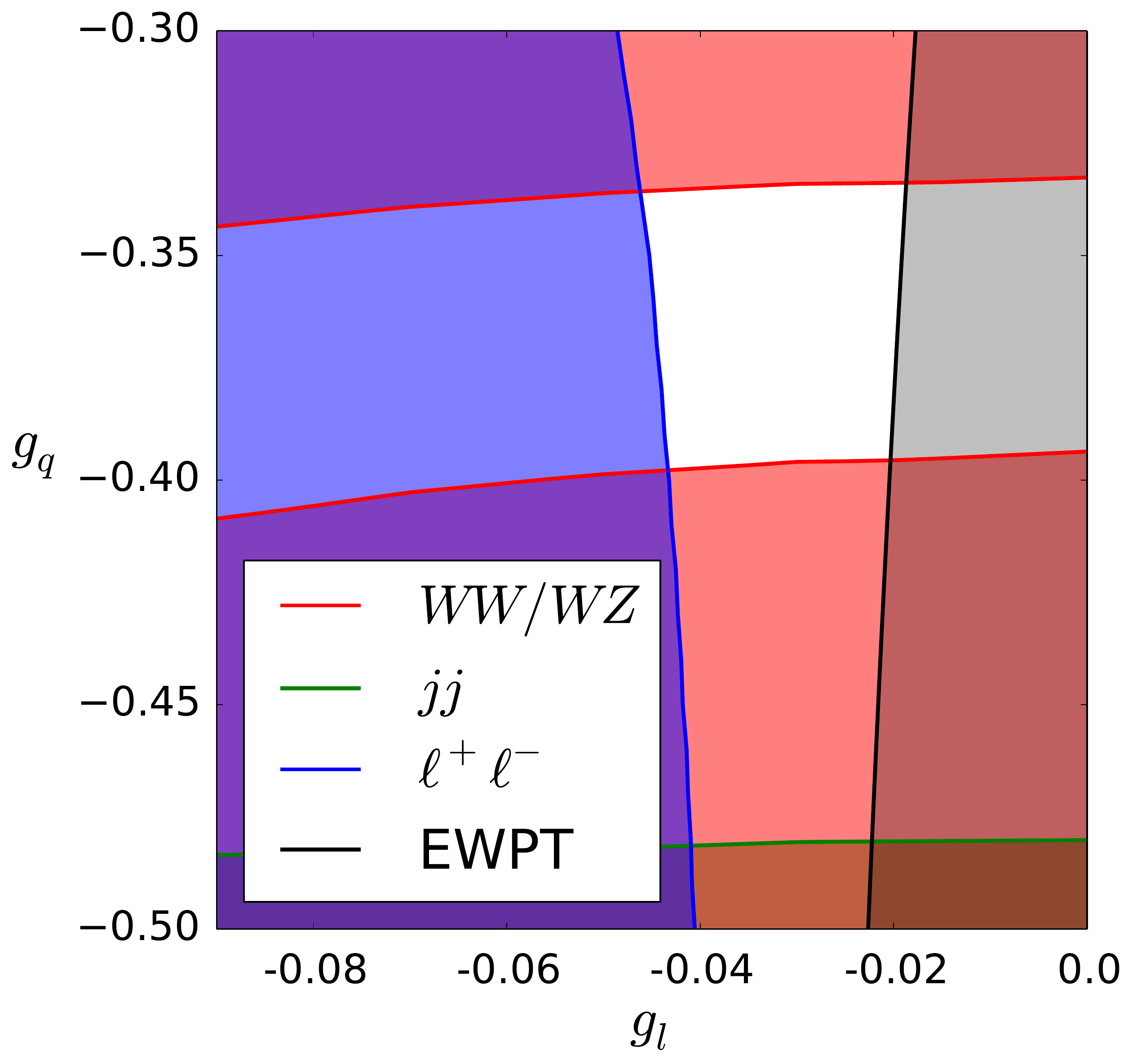}
&
\includegraphics[width=0.4\textwidth]{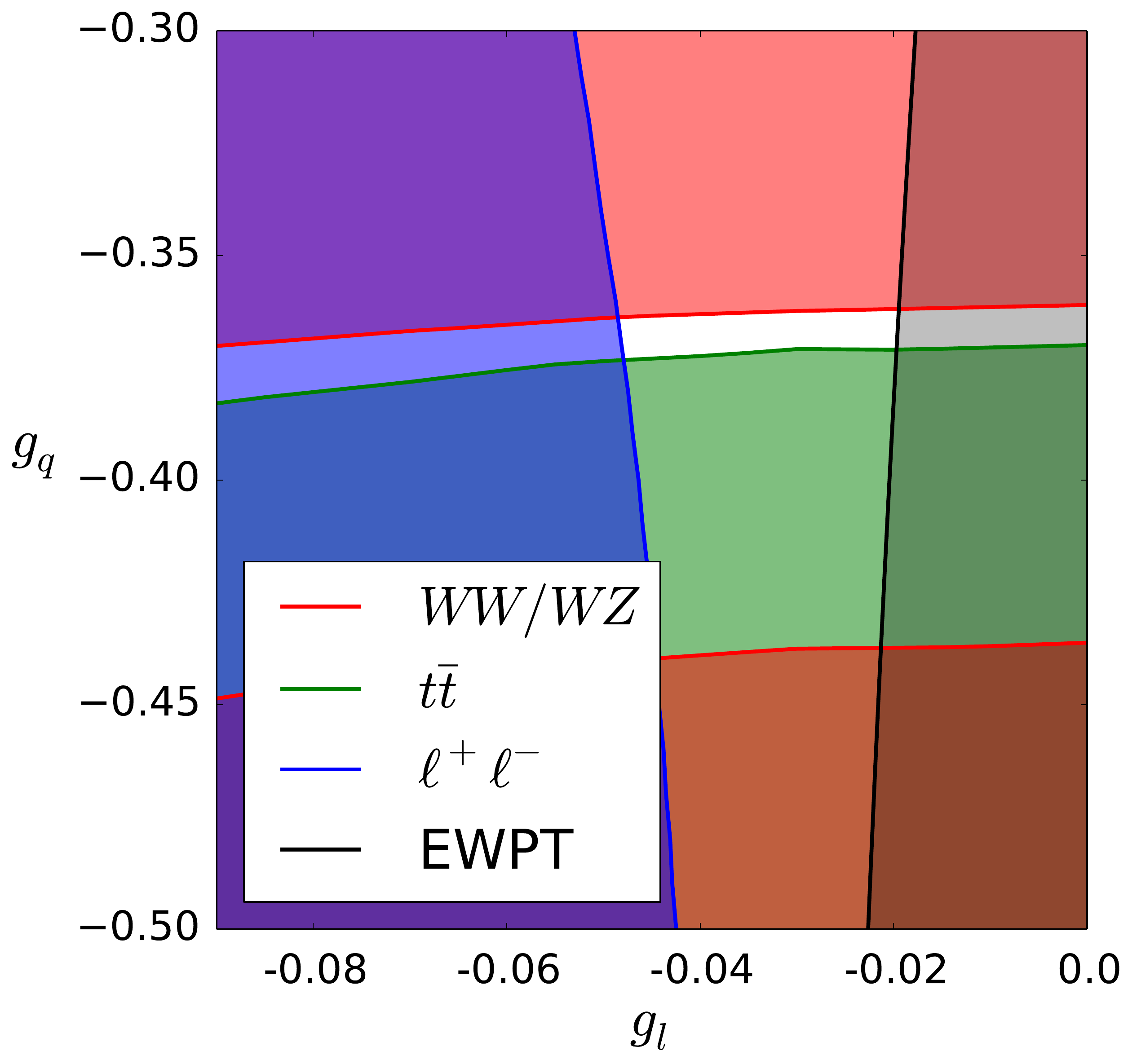}
\end{tabular}
\caption{Constraints on the model from the diboson signal, $t\bar{t}$, 
  $jj$, $\ell^+ \ell^-$ and EWPT. The colored areas correspond
  to the region excluded by the different bounds.
The input parameters in the figure have been fixed to $M=1.8$ TeV and
$g_V=0.75$, $M=1.9$ TeV and $g_V=0.8$ 
and $M=2$ TeV and $g_V=0.85$ 
in the top, middle and bottom rows, respectively. The
left (right) column corresponds to $g_{t_R}=0.3$ (0.5) and in all six
plots we have fixed $g_{q_3}=0.3$.
\label{results:fig}
}
\end{figure}

We have performed detailed scans over the parameter space of the
model, computing the constraints from EWPT and the
corresponding production cross sections in all the relevant channels
discussed in the previous sections. These
cross sections have been computed with
\texttt{MG5}~\cite{Alwall:2014hca}, 
using a model created with FeynRules~\cite{Alloul:2013bka}.
Our results are summarized in Fig.~\ref{results:fig} in which we show
the region in the 
$g_l-g_q$ that successfully reproduces the ATLAS excess, together with
the most relevant constraints. The colored areas correspond
to the regions that are forbidden by current constraints. In the case
of the $WW/WZ$ signal, this means that either the sum of the cross
section in both channels is smaller than $5$ fb (upper red region) or
that the bound in Eq.~(\ref{diboson:upper:bound}) is not satisfied
(lower red region). All the other bounds are color coded as the label
indicates. The white area corresponds to the allowed region that generates the
observed signal excess.
The input parameters in the figure have been fixed to $M=1.8$ TeV and
$g_V=0.75$, $M=1.9$ TeV and $g_V=0.8$ 
and $M=2$ TeV and $g_V=0.85$ 
in the top, middle and bottom rows, respectively. The
left (right) column corresponds to $g_{t_R}=0.3$ (0.5) and in all six
plots we have fixed $g_{q_3}=0.3$.

These plots show that there is a very narrow range in $g_l$ that is
compatible with EWPT and dilepton bounds. Having enough signal
requires sizeable $g_V$ and $g_q$. Bounds from $t\bar{t}$
production constrain the value of $g_{t_R}$ and $g_{q_3}$ (the latter
is also constrained by EWPT) to be relatively low. Finally, when the
coupling to the third generation quarks are small enough, the upper
bound on diboson production in Eq.~(\ref{diboson:upper:bound}) and dijet
production become the
most stringent constraints that prevent the
strength of $g_q$ to be too large. Combining all these constraints we
find the following region in the parameter space for our non-custodial
composite Higgs model that can explain the observed diboson excess
without contradicting any other experimental bound
\begin{equation}
g_V\sim 0.8-0.9, \quad g_{t_R} \sim g_{q_3}\sim 0.3,
\quad -0.05 \lesssim g_l \lesssim -0.01,
\quad -0.5 \lesssim g_{q} \lesssim -0.3,
\end{equation}
where, depending on the exact values of the parameters (including $M$)
slightly smaller or larger ranges can be allowed.

Before going to our conclusions let us mention that there are some
small excesses in other analyses, like $WH$ or dijets which could also
be explained, at least partially, in the same region of parameter
space needed to explain the resonant diboson excess. We have not tried
to fit any of them but it is worth keeping an eye on any forthcoming
analysis. Any complete UV model will have different correlated
predictions in different channels that hopefully the run 2 of the LHC
will be able to confirm. 

\section{Conclusions\label{conclu}}

In this paper we have tried to give an explanation to the recently
reported excess in diboson resonant production in the context of
non-custodial composite models. 
Composite Higgs models 
naturally predict the existence of spin-1 resonances, among other
ones, which could explain the observed excess. We have
shown that in a particular class of model based on a soft-wall
construction one can satisfy all bounds from EWPT and direct searches
and still be able to reproduce the resulting observed excess. 
The mass range reported for the excess, $M\sim 2$ TeV, is
however somewhat low to be accommodated in standard warped models, 
due to the strong bound coming from the $S$-parameter. This is the
reason to resort to soft-wall models, in which the mixing of these
resonances with the SM gauge bosons is reduced in such a way that
$\sim 2$ TeV vector resonances are compatible with bounds from EWPT.
For the sake of simplicity we have considered a minimal non-custodial
composite Higgs model. However we have prefered to remain agnostic
regarding the underlying gravitational model and we have therefore
parameterized with complete generality the relevant couplings. Despite
the tension present in the model between the production of a sizeable
diboson cross section and direct and indirect constraints we have
found that there are regions of parameter space in which the excess
can be completely explained without contradicting any other
experimental bound. 
The global analysis, including all the constraints, that we have
performed selects a well-defined region of parameter space in
non-custodial composite Higgs models in which lepton-quark
universality have to be abandoned (due to the bounds from $\ell^+ \ell^-$
searches). We also need a sizeable coupling to the light quarks and a
reduced coupling to third generation quarks. This would generically
correspond to a holographic soft-wall model with a modest volume
factor, essentially conformal leptons and with a not-fully composite
top quark.
The tension between the signal and experimental bounds also means that, if
the excess is confirmed during the LHC run 2, it is likely that other
related excesses appear with a similar mass scale. In particular,
$t\bar{t}$, dijet, $\ell^+ \ell^-$ or diboson resonances in other
channels (most notably in the semi-leptonic ones) are expected to be
easily visible with the new data. A detailed measurement of the
corresponding signals would help disentangle the precise region of
parameter space that explains all the excesses.

\section*{Acknowledgments}

We would like to thank Jorge de Blas and Gustaaf Brooijmans 
for useful discussions and the  \'Ecole de  Physique des Houches for
creating an ideal atmosphere where this project was started. AC is
supported by the Swiss National Science Foundation under 
contract SNSF 200021-143781. The work of AD was supported
in part by the National Science Foundation under Grant
No. PHY-1215979. MQ~is partly supported by the Spanish 
Consolider-Ingenio 2010 Programme CPAN (Grant number CSD2007-00042), by
Grant number CICYT-FEDER-FPA2011-25948, by the Severo Ochoa excellence
program of 
MINECO under Grant number SO-2012-0234 and by Secretaria d'Universitats i
Recerca del Departament d'Economia i Coneixement de la Generalitat de
Catalunya under Grant number2014 SGR 1450. The work of JS has been
partially supported by the European Commission (PITN-GA-2012-316704
HIGGSTOOLS), by MINECO under grants 
(FPA2010-17915 and FPA2013-47836-C3-2-P) and by Junta de
Andaluc\'{\i}a grants FQM 101 and FQM 6552.




\begin{thebibliography}{99}

\bibitem{Aad:2015owa}
  G.~Aad {\it et al.}  [ATLAS Collaboration],
  arXiv:1506.00962 [hep-ex].

\bibitem{Khachatryan:2014hpa}
  V.~Khachatryan {\it et al.}  [CMS Collaboration],
  JHEP {\bf 1408} (2014) 173
  [arXiv:1405.1994 [hep-ex]].


\bibitem{Fukano:2015hga}
  H.~S.~Fukano, M.~Kurachi, S.~Matsuzaki, K.~Terashi and K.~Yamawaki,
  arXiv:1506.03751 [hep-ph];
  J.~Hisano, N.~Nagata and Y.~Omura,
  arXiv:1506.03931 [hep-ph];
  D.~B.~Franzosi, M.~T.~Frandsen and F.~Sannino,
  arXiv:1506.04392 [hep-ph];
  K.~Cheung, W.~Y.~Keung, P.~Y.~Tseng and T.~C.~Yuan,
  arXiv:1506.06064 [hep-ph];
  S.~S.~Xue,
  arXiv:1506.05994 [hep-ph];
  B.~A.~Dobrescu and Z.~Liu,
  arXiv:1506.06736 [hep-ph];
  J.~A.~Aguilar-Saavedra,
  arXiv:1506.06739 [hep-ph];
  A.~Alves, A.~Berlin, S.~Profumo and F.~S.~Queiroz,
  arXiv:1506.06767 [hep-ph];
  Y.~Gao, T.~Ghosh, K.~Sinha and J.~H.~Yu,
  arXiv:1506.07511 [hep-ph];
  A.~Thamm, R.~Torre and A.~Wulzer,
  arXiv:1506.08688 [hep-ph];
  J.~Brehmer, J.~Hewett, J.~Kopp, T.~Rizzo and J.~Tattersall,
  arXiv:1507.00013 [hep-ph];
  Q.~H.~Cao, B.~Yan and D.~M.~Zhang,
  arXiv:1507.00268 [hep-ph];
  G.~Cacciapaglia and M.~T.~Frandsen,
  arXiv:1507.00900 [hep-ph].



\bibitem{Agashe:2003zs}
  K.~Agashe, A.~Delgado, M.~J.~May and R.~Sundrum,
  JHEP {\bf 0308} (2003) 050
  [hep-ph/0308036];
  K.~Agashe, R.~Contino and A.~Pomarol,
  Nucl.\ Phys.\ B {\bf 719} (2005) 165
  [hep-ph/0412089];
  M.~Carena, E.~Ponton, J.~Santiago and C.~E.~M.~Wagner,
  Nucl.\ Phys.\ B {\bf 759} (2006) 202
  [hep-ph/0607106];
  Phys.\ Rev.\ D {\bf 76} (2007) 035006
  [hep-ph/0701055].

\bibitem{Falkowski:2008fz}
  A.~Falkowski and M.~Perez-Victoria,
  JHEP {\bf 0812} (2008) 107
  [arXiv:0806.1737 [hep-ph]].

\bibitem{Cabrer:2010si}
  J.~A.~Cabrer, G.~von Gersdorff and M.~Quiros,
  Phys.\ Lett.\ B {\bf 697} (2011) 208
  [arXiv:1011.2205 [hep-ph]];
  JHEP {\bf 1105} (2011) 083
  [arXiv:1103.1388 [hep-ph]].

\bibitem{Carmona:2011ib}
  A.~Carmona, E.~Ponton and J.~Santiago,
  JHEP {\bf 1110} (2011) 137
  [arXiv:1107.1500 [hep-ph]].
\bibitem{deBlas:2012qf}
  J.~de Blas, A.~Delgado, B.~Ostdiek and A.~de la Puente,
  Phys.\ Rev.\ D {\bf 86} (2012) 015028
  [arXiv:1206.0699 [hep-ph]].


\bibitem{Aad:2015ufa}
  G.~Aad {\it et al.}  [ATLAS Collaboration],
  Eur.\ Phys.\ J.\ C {\bf 75} (2015) 5,  209
  [arXiv:1503.04677 [hep-ex]].

\bibitem{Khachatryan:2014gha}
  V.~Khachatryan {\it et al.}  [CMS Collaboration],
  JHEP {\bf 1408} (2014) 174
  [arXiv:1405.3447 [hep-ex]].



\bibitem{Aad:2014xka}
  G.~Aad {\it et al.}  [ATLAS Collaboration],
  Eur.\ Phys.\ J.\ C {\bf 75} (2015) 2,  69
  [arXiv:1409.6190 [hep-ex]].

\bibitem{Aad:2014pha}
  G.~Aad {\it et al.}  [ATLAS Collaboration],
  Phys.\ Lett.\ B {\bf 737} (2014) 223
  [arXiv:1406.4456 [hep-ex]].

\bibitem{Khachatryan:2014xja}
  V.~Khachatryan {\it et al.}  [CMS Collaboration],
  Phys.\ Lett.\ B {\bf 740} (2015) 83
  [arXiv:1407.3476 [hep-ex]].

\bibitem{Aad:2015yza}
  G.~Aad {\it et al.}  [ATLAS Collaboration],
  Eur.\ Phys.\ J.\ C {\bf 75} (2015) 6,  263
  [arXiv:1503.08089 [hep-ex]].

\bibitem{Khachatryan:2015bma}
  V.~Khachatryan {\it et al.}  [CMS Collaboration],
  arXiv:1506.01443 [hep-ex].

\bibitem{Aad:2014cka}
  G.~Aad {\it et al.}  [ATLAS Collaboration],
  Phys.\ Rev.\ D {\bf 90} (2014) 5,  052005
  [arXiv:1405.4123 [hep-ex]].

\bibitem{ATLAS:2014wra}
  G.~Aad {\it et al.}  [ATLAS Collaboration],
  JHEP {\bf 1409} (2014) 037
  [arXiv:1407.7494 [hep-ex]].
  
\bibitem{Chatrchyan:2012oaa}
  S.~Chatrchyan {\it et al.}  [CMS Collaboration],
  Phys.\ Lett.\ B {\bf 720} (2013) 63
  [arXiv:1212.6175 [hep-ex]].
  
\bibitem{Khachatryan:2014tva}
  V.~Khachatryan {\it et al.}  [CMS Collaboration],
  Phys.\ Rev.\ D {\bf 91} (2015) 9,  092005
  [arXiv:1408.2745 [hep-ex]].
  
\bibitem{Aad:2015fna}
  G.~Aad {\it et al.}  [ATLAS Collaboration],
  arXiv:1505.07018 [hep-ex].
  
\bibitem{Khachatryan:2015sma}
  V.~Khachatryan {\it et al.}  [CMS Collaboration],
  arXiv:1506.03062 [hep-ex].
  
\bibitem{Aad:2014aqa}
  G.~Aad {\it et al.}  [ATLAS Collaboration],
  Phys.\ Rev.\ D {\bf 91} (2015) 5,  052007
  [arXiv:1407.1376 [hep-ex]].
  
\bibitem{Khachatryan:2015sja}
  V.~Khachatryan {\it et al.}  [CMS Collaboration],
  Phys.\ Rev.\ D {\bf 91} (2015) 5,  052009
  [arXiv:1501.04198 [hep-ex]].
  

\bibitem{twiki}
\texttt{https://atlas.web.cern.ch/Atlas/GROUPS/PHYSICS/PAPERS/EXOT-2013-08/}

\bibitem{Kaplan:1983fs}
  D.~B.~Kaplan and H.~Georgi,
  Phys.\ Lett.\ B {\bf 136} (1984) 183;
  D.~B.~Kaplan, H.~Georgi and S.~Dimopoulos,
  Phys.\ Lett.\ B {\bf 136} (1984) 187;
  R.~Contino, Y.~Nomura and A.~Pomarol,
  Nucl.\ Phys.\ B {\bf 671} (2003) 148
  [hep-ph/0306259].



\bibitem{Barcelo:2011wu}
  R.~Barcelo, A.~Carmona, M.~Chala, M.~Masip and J.~Santiago,
  Nucl.\ Phys.\ B {\bf 857} (2012) 172
  [arXiv:1110.5914 [hep-ph]];
  C.~Bini, R.~Contino and N.~Vignaroli,
  JHEP {\bf 1201} (2012) 157
  [arXiv:1110.6058 [hep-ph]];
  M.~Chala, J.~Juknevich, G.~Perez and J.~Santiago,
  JHEP {\bf 1501} (2015) 092
  [arXiv:1411.1771 [hep-ph]];
  A.~Azatov, D.~Chowdhury, D.~Ghosh and T.~S.~Ray,
  arXiv:1505.01506 [hep-ph].

\bibitem{delAguila:2010vg}
  F.~del Aguila, A.~Carmona and J.~Santiago,
  JHEP {\bf 1008} (2010) 127
  [arXiv:1001.5151 [hep-ph]];
  J.~A.~Cabrer, G.~von Gersdorff and M.~Quiros,
  JHEP {\bf 1201} (2012) 033
  [arXiv:1110.3324 [hep-ph]];
  G.~von Gersdorff, M.~Quiros and M.~Wiechers,
  JHEP {\bf 1302} (2013) 079
  [arXiv:1208.4300 [hep-ph]];
  A.~Carmona and F.~Goertz,
  JHEP {\bf 1505} (2015) 002
  [arXiv:1410.8555 [hep-ph]].


\bibitem{Davoudiasl:2009cd}
  H.~Davoudiasl, S.~Gopalakrishna, E.~Ponton and J.~Santiago,
  New J.\ Phys.\  {\bf 12} (2010) 075011
  [arXiv:0908.1968 [hep-ph]].

\bibitem{Efrati:2015eaa}
  A.~Efrati, A.~Falkowski and Y.~Soreq,
  arXiv:1503.07872 [hep-ph].

\bibitem{deBlas:2013qqa}
  J.~de Blas, M.~Chala and J.~Santiago,
  Phys.\ Rev.\ D {\bf 88} (2013) 095011
  [arXiv:1307.5068 [hep-ph]].


\bibitem{Anastasiou:2009rv}
  C.~Anastasiou, E.~Furlan and J.~Santiago,
  Phys.\ Rev.\ D {\bf 79} (2009) 075003
  [arXiv:0901.2117 [hep-ph]];
  C.~Grojean, O.~Matsedonskyi and G.~Panico,
  JHEP {\bf 1310} (2013) 160
  [arXiv:1306.4655 [hep-ph]];
  R.~Contino and M.~Salvarezza,
  arXiv:1504.02750 [hep-ph].

\bibitem{Alwall:2014hca}
  J.~Alwall, R.~Frederix, S.~Frixione, V.~Hirschi, F.~Maltoni, O.~Mattelaer, H.-S.~Shao and T.~Stelzer {\it et al.},
  JHEP {\bf 1407} (2014) 079
  [arXiv:1405.0301 [hep-ph]].

\bibitem{Alloul:2013bka}
  A.~Alloul, N.~D.~Christensen, C.~Degrande, C.~Duhr and B.~Fuks,
  Comput.\ Phys.\ Commun.\  {\bf 185} (2014) 2250
  [arXiv:1310.1921 [hep-ph]].

\end{thebibliography}
\end{document}